\RequirePackage[final]{graphicx}
\documentclass[draft,preprint,3p,twocolumn]{elsarticle}

\usepackage[T1]{fontenc}
\usepackage[utf8]{inputenc}
  
\usepackage{ifdraft}
\usepackage{amssymb}
\usepackage{lineno,hyperref}
\usepackage[plain]{fancyref}
\usepackage{xspace}
\usepackage{acronym}
\usepackage{booktabs}
\usepackage{multirow}
\usepackage{relsize}
\usepackage{pdflscape}
\usepackage{etoolbox}
\usepackage{rotating}
\usepackage{lscape}
\usepackage[inline]{enumitem}
\usepackage{tikz}
\usetikzlibrary{trees}
\usetikzlibrary{fit}
\usetikzlibrary{calc}
\usepackage{tikz-qtree}
\usepackage[final]{listings}
\usepackage{caption}
\usepackage{subcaption}
\usepackage{ifthen}
\usepackage{url}
\usepackage{xcolor}

\usepackage{algorithmicx}
\usepackage{algpseudocode}
\usepackage{algorithm}
\usepackage{verbatim}

\definecolor{author}{rgb}{.0, .0, .0}
\definecolor{comment}{rgb}{.1, .0, .9}
\definecolor{note}{rgb}{.9, .4, .0}
\definecolor{idea}{rgb}{.1, .7, .0}
\definecolor{missing}{rgb}{.9, .1, .0}

\algnewcommand{\LeftComment}[1]{\Statex \(\triangleright\) #1}

\providecommand{\comment}[2]{}

\def\fref{\Fref} 
\renewcommand{\lstlistingname}{Snippet}
\newcommand*{\fancyreflstlabelprefix}{lst}
\newcommand*{\Freflstname}{\lstlistingname}
\newcommand*{\freflstname}{\MakeLowercase{\lstlistingname}}
\Frefformat{vario}{\fancyreflstlabelprefix}%
  {\Freflstname\fancyrefdefaultspacing#1#3}
\frefformat{vario}{\fancyreflstlabelprefix}%
  {\freflstname\fancyrefdefaultspacing#1#3}
\Frefformat{plain}{\fancyreflstlabelprefix}%
  {\Freflstname\fancyrefdefaultspacing#1}
\frefformat{plain}{\fancyreflstlabelprefix}%
  {\freflstname\fancyrefdefaultspacing#1}

\newcommand*{\fancyrefalglabelprefix}{alg} 
\newcommand*{\Frefalgname}{Algorithm}%
\newcommand*{\frefalgname}{%
 \MakeLowercase{\Frefalgname}}%
\Frefformat{vario}{\fancyrefalglabelprefix}%
  {\Frefalgname\fancyrefdefaultspacing#1#3}
\frefformat{vario}{\fancyrefalglabelprefix}%
  {\frefalgname\fancyrefdefaultspacing#1#3}
\Frefformat{plain}{\fancyrefalglabelprefix}%
  {\Frefalgname\fancyrefdefaultspacing#1}
\frefformat{plain}{\fancyrefalglabelprefix}%
  {\frefalgname\fancyrefdefaultspacing#1}
  
\newcommand*{\fancyreflnlabelprefix}{ln}
\newcommand*{\Freflnname}{Line}
\newcommand*{\freflnname}{\MakeLowercase{\Freflnname}}
\Frefformat{vario}{\fancyreflnlabelprefix}%
  {\Freflnname\fancyrefdefaultspacing#1#3}
\frefformat{vario}{\fancyreflnlabelprefix}%
  {\freflnname\fancyrefdefaultspacing#1#3}
\Frefformat{plain}{\fancyreflnlabelprefix}%
  {\Freflnname\fancyrefdefaultspacing#1}
\frefformat{plain}{\fancyreflnlabelprefix}%
  {\freflnname\fancyrefdefaultspacing#1}

\renewcommand{\lstlistingname}{Snippet}

\lstdefinestyle{floating}{%
  float=htbp,
  captionpos=b,
  numberstyle=\scriptsize\color{gray},
  numbersep=2pt,
  belowskip=-0\baselineskip,
  aboveskip=-0\baselineskip,
  escapechar=`
}

\lstdefinelanguage{JavaScript}{
keywords={},
keywordstyle=\color{purple}\bfseries,
ndkeywords={typeof, new, true, false, catch, function, return, null, catch, switch, var, if, in, for, while, do, else, case, break, throw, this, instanceof},
ndkeywordstyle=\bfseries,
identifierstyle=\color{black},
sensitive=false,
comment=[l]{//},
morecomment=[s]{/*}{*/},
stringstyle=\color{green!50!brown}\ttfamily,
morestring=[b]',
morestring=[b]"
}

\lstdefinestyle{ctxtraits}
 {language=JavaScript,
  frame=lines,
  showstringspaces=false,
  tabsize=3,
  style=floating,
  morekeywords={Trait, cop, Context, activate, deactivate, adapt, addObjectPolicy, manager}
}

\lstnewenvironment{ctxtraits}[1][]
 {\lstset{style=ctxtraits,#1}}{}  
 
\newcommand{\scode}[1]{\textrm{\texttt{#1}}}
\def\scode{\lstinline[style=ctxtraits]}

\lstnewenvironment{console}[1][]
 {\renewcommand{\lstlistingname}{Console output}%
  \lstset{numbers=left,basicstyle=\small\ttfamily,#1}}{}

\newcommand{\ctx}[1]{\texttt{\textsc{#1}}}  

\newcommand{\specialterm}[2][]%
  {\textit{#2}\ifdraft{\ifblank{#1}{}{\marginpar{\sffamily\scriptsize#1}}}{}}

\newcommand{\eg}{\emph{e.g.,}\xspace}
\newcommand{\ie}{\emph{i.e.,}\xspace}

\acrodef{API}{Application Programming Interface}
\acrodef{ADT}{Abstract Data Type}
\acrodef{AOP}{Aspect-oriented Programming}
\acrodef{COP}{Context-oriented Programming}
\acrodef{SPL}{Software Product Line}
\acrodef{DSL}{Domain-Specific Language}
\acrodef{FOP}{Feature-oriented Programming}
\acrodef{IOT}[IoT]{Internet of Things}
\acrodef{ML}{Machine Learning}
\acrodef{LOC}{Lines of Code}
\acrodef{MOP}{metaobject protocol}
\acrodef{RL}{Reinforcement Learning}
\acrodef{STM}{Software Transactional Memory}
\acrodef{CPS}{Cyber Physical System}
\acused{API} 

\journal{Journal of Information \& Software Technology}

\begin{document}

\begin{frontmatter}
\title{Auto-COP: Adaptation Generation in Context-Oriented Programming Using Reinforcement Learning Options}


\author{Nicol\'as Cardozo\corref{cor1}} 
\address{Systems and Computing Engineering Department, Universidad de los Andes. Bogot\'a, Colombia.}
\ead{n.cardozo@uniandes.edu.co}

\author{Ivana Dusparic} 
\address{School of Computer Science and Statistics, Trinity College Dublin. Ireland.}
\ead{ivana.dusparic@scss.tcd.ie}

\cortext[cor1]{Corresponding author} 


\begin{abstract}
\textbf{Context:} Self-adaptive software systems continuously adapt in response to 
internal and external changes in their execution environment, captured 
as contexts. The \ac{COP} paradigm posits a technique for the 
development of self-adaptive systems, capturing their main 
characteristics with specialized programming language constructs. In 
\ac{COP}, adaptations are specified as independent modules that are 
composed in and out of the base system as contexts are activated and 
deactivated in response to sensed circumstances from the surrounding 
environment. However, the definition of adaptations, their contexts and 
associated specialized behavior, need to be specified at design time. 
In complex cyber physical systems this is intractable, if not impossible, due to new unpredicted operating conditions arising. 
\textbf{Objective:} In 
this paper, we propose Auto-\ac{COP}, a new technique to enable 
generation of adaptations at run time. Auto-\ac{COP} uses \ac{RL} 
options to build action sequences, based on the previous instances of 
the system execution (for example, atomic system actions enacted by human operators). 
Options are further explored in interaction with the environment, and 
the most suitable options for each context are used to generate the  
adaptations, exploiting \ac{COP} abstractions. 
\textbf{Method:} To validate Auto-\ac{COP}, we present two case 
studies exhibiting different system characteristics and application 
domains: a driving assistant and a robot delivery system. We present 
examples of Auto-\ac{COP} to illustrate the types of circumstances 
(contexts) requiring adaptation at run time, and the 
corresponding generated adaptations for each context. 
\textbf{Results:} We confirm that the generated adaptations exhibit 
correct system behavior measured by domain-specific performance 
metrics (\eg conformance to specified speed limit), while reducing 
the number of required execution/actuation steps by a factor of two 
showing that the adaptations are regularly selected by the running 
system as adaptive behavior is more appropriate than the execution of 
atomic actions. \textbf{Conclusion:} Therefore, we 
demonstrate that Auto-\ac{COP} is able to increase system adaptivity 
by enabling run-time generation of new adaptations for conditions 
detected at run time, while retaining the modularity offered by 
\ac{COP} languages, and reducing the upfront specification required by system developers. 
\end{abstract}


\end{frontmatter}

\acresetall


\section{Introduction}
\label{sec:introduction}

Self-adaptive systems~\cite{salehie09adaptive} gather information from
the execution environment (both internal and external) through 
monitors and sensors. Such information is used to dynamically adapt 
the system's behavior, as a means to offer a more appropriate behavior for the 
situation at hand. The \ac{COP} paradigm~\cite{hirschfeld+08contexts} 
addresses self-adaptation at the programming language level, to enable 
a clean modularization of adaptations from the base program logic as 
well as from other adaptations. 

Three main concepts are behind dynamic adaptations in \ac{COP}: 
\emph{contexts}, \emph{behavioral variations}, and 
\emph{context activations}. 
Contexts are defined as first class entities of the system that 
capture meaningful situations from the system's surrounding execution 
environment. Behavioral variations realize modular partial behavior 
(\eg a method) specifications defined in isolation of other 
components. Context activations take place whenever the situations 
they represent are sensed in the environment, composing their 
associated behavioral adaptation into the system. In this manner, the 
dynamic composition model used in \ac{COP} reifies the MAPE 
(Monitor-Analyze-Plan-Execute) loop~\cite{brun09}.

\ac{COP} has been used to model self-adaptive systems in 
general~\cite{cardozo20ecas, cardozo21}. Concretely, \ac{COP} 
\emph{Monitors} specified system variables or the surrounding environment in the 
\emph{Context Discovery} (internal) module~\cite{cardozo13phd}. Monitored variables are 
used for context activation. Taking into account the values for variables, the \emph{Context Manager} 
determines whether to activate or deactivate a context; serving as the \emph{Analysis} phase 
of the MAPE. In \ac{COP}, adaptation \emph{Planning} is implicit and managed by the language, as 
adaptations are composed with the system behavior as soon as possible~\cite{cardozo+11ctxact}. 
Adaptation \emph{Execution} is 
coupled with the planning, as adaptations are enacted whenever contexts are activated, or upon 
their use. Finally, note that the \emph{Knowledge} component is only present implicitly in \ac{COP}. 
The knowledge for adaptations is encoded within the aforementioned modules, as the system 
uses as a precondition the fact that adaptations are appropriate to their context, and the context is 
sensed from the environment.

\ac{COP} has proven effective in achieving dynamic adaptations~\cite{appeltauer11, salvaneschi11, kamina14, watanabe15, duhoux19eics, cardozo21, leger23}, 
although latent problems arise from the definition of adaptations.  
The behavior of adaptations still needs to be defined 
beforehand by developers. This is a challenging task, and in many 
cases not viable, as the execution environment may be unknown. 

To address the problem of upfront definition of adaptations, in this 
paper we propose Auto-\ac{COP}, a new technique to further system 
automation by enabling the dynamic generation of required adaptations 
in \ac{COP}. Auto-\ac{COP} uses \ac{RL} options~\cite{sutton18} to 
build action sequences, based on the previous instances of the system 
execution (for example, actions enacted by system users).  An \ac{RL} 
option encodes sequences of actions, and is defined by a set of states in 
which it can be initiated (\ie contexts in \ac{COP} terminology), an internal 
policy (\ie sequence of actions to take), and a terminating condition 
(\ie context deactivation). Through 
options, Auto-\ac{COP} identifies sequences of actions (referred to as 
\emph{atomic actions} in RL options terminology, the term which we 
also adopt in this paper) and the environmental conditions in which 
these take place, extrapolating them to \emph{behavioral variations} 
and their associated \emph{context}, to realize dynamic adaptations 
as in \ac{COP}. As a consequence, we generate adaptations from learned 
options, that can be autonomously executed whenever their context is 
sensed. Note that action sequences executed for a particular situation 
(\eg program state) can differ during execution. By using options, 
Auto-\ac{COP} can differentiate between the executed atomic action 
sequences, and via interactions with the environment learn the option 
that is currently most appropriate for the system execution. 

\begin{figure}[htbp]
	\centering
	\includegraphics[width=\linewidth]{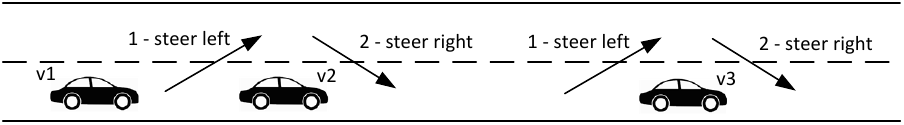}
	\caption{Driving assistant overtaking scenario}
	\label{fig:overtake}
\end{figure}

To illustrate the problem and our approach, we consider the case 
of a driving assistant system (further discussed in 
\fref{sec:driving-assistant}). Vehicles 
are controlled by five atomic intervention actions: 
\scode{straight}, \scode{steer right}, \scode{steer left}, 
\scode{speed up}, and \scode{slow down}. The driving process 
consists of a series of actions, which are frequently repeated in 
response to certain conditions on the road. For example, in 
\fref{fig:overtake}, when vehicle $v1$ encounters $v2$ driving at a 
lower speed lower than its own, it performs actions \scode{steer left} 
and \scode{steer right} in order to overtake it. When it encounters 
the next vehicle, $v3$, the same actions \scode{steer left} and 
\scode{steer right} are repeated. Such ordered sequences of 
actions are precisely the kind of behavior that Auto-\ac{COP} could 
identify, group, extract, and generate \ac{COP}-based adaptations to 
automate behavior as an enhancement to the driving assistant system. 
	
In this example, the base system behavior is to drive straight on the 
right-lane of the road. To keep appropriate behavior of the system 
under different situations we must adapt its behavior, executing 
different behavior for particular situations. Such situations are 
defined as execution contexts. For example, on the left-hand side of 
\fref{fig:overtake} $v1$ is in a special situation, with a close 
proximity to $v2$. This is defined as a context. Associated to the 
context, we can define a \emph{behavioral adaptation}, a 
specialization of the base behavior, in our case, consisting of the 
\scode{steer left} and \scode{steer right} behavior. The combination 
of the context and its associated behavioral adaptation constitute an 
adaptation. Once defined, every time the system senses the context, 
for example by means of external sensors, it triggers the context 
activation, which effectively composes the behavior associated with 
this context with the application. Whenever the context is no longer 
sensed, it is deactivated, effectively withdrawing the associated 
behavioral adaptations from the application, and reverting it to its 
base behavior.

While our motivating example is simple, in complex cyber physical 
systems such sequences might have dozens or hundreds of steps. For 
example in Minecraft, a complex game environment which is extensively 
used for novel \ac{RL} techniques benchmarking, \ac{RL} options were 
successfully used to build action sequences and reduce 
the number of steps in a trajectory 100-fold, from 10,000 steps to only
100~\cite{minecraft2020}. 

The main advantage of Auto-\ac{COP} is that it allows adaptivity 
of self-adaptive systems, by enabling run-time 
generation of new adaptations using RL, while retaining modularity and 
remaining compatible with existing \ac{COP} tools and techniques. 
Using Auto-\ac{COP}, new behavior that corresponds to newly arisen 
contexts, not foreseen at design time, do not require software 
developers to specify and incorporate them into the system. Rather, 
Auto-\ac{COP} is the first approach to enable building \ac{COP} adaptations on-line from actions executed during user 
interactions or operator interventions at run time. In such a way, 
the system can learn directly from human users and/or operators, 
removing the need for manual intervention when the given context 
arises in the future. In addition, if previously defined or 
generated adaptations are not suitable to respond to a given 
context any longer, new ones will be generated and overwrite old 
unsuitable behavior, without the involvement of developers. While 
"pure" RL-based techniques have been used extensively in 
self-adaptive systems to achieve run-time adaptation~\cite{dangelo19}, 
such approaches are fully off-line and still require upfront 
definition of adaptations~\cite{jamshidi19}. Moreover, existing 
approaches in the literature are application-specific and lack the 
modularity and reusability offered by combining RL with \ac{COP}.

To validate the effectiveness of Auto-\ac{COP} to generate adaptations 
at run time that can be integrated with a \ac{COP} system to drive 
dynamic adaptations, we 
apply our technique to two different application domains: a driving 
assistant and a warehouse robot delivery system. Our evaluation in 
each case study focuses on two main aspects: verifying that 
adaptations are automatically generated and used by the system in 
subsequent executions, as well as that generated adaptations exhibit 
correct system behavior with respect to system goals. We observe that 
the system executing the adaptations maintains or improves the 
performance of the base system executing atomic actions. 
Furthermore, we observe twofold reduction in required 
execution/actuation steps to achieve the same tasks. This indicates 
that the generated options are indeed preferred by the system and that 
a single step may trigger the execution of multiple actions (a learned 
sequence) rather than atomic actions. We conclude that 
Auto-\ac{COP} is able to increase system adaptivity 
by enabling run-time generation of new adaptations for conditions 
detected at run time, while retaining modularity of \ac{COP}. 

In summary, the main contributions of our paper are:
\begin{itemize}[leftmargin=10pt]
  \item Auto-\ac{COP}, a novel approach  which utilizes \ac{RL} options to generate adaptations at run time, based on previous system executions and interaction with the environment. Adaptations are generated as defined in \ac{COP}, and as such can be integrated into \ac{COP}-based systems to enable dynamic self-adaptation at run time.
  \item We detail the steps required to incorporate Auto-\ac{COP} in a self-adaptive system, providing the specifics of the algorithm and code snippets.
  \item We apply Auto-\ac{COP} in two separate application domains and confirm its ability to generate adaptations autonomously, to be dynamically incorporated to, and withdrawn from the system.
\end{itemize}

The rest of the paper is structured as follows. \fref{sec:background} introduces necessary concepts from \ac{COP} and \ac{RL} options.   \fref{sec:learning-adaptations} presents the details of Auto-\ac{COP} approach, while its evaluation in two case studies is presented in \fref{sec:validation}. In \fref{sec:related} we discuss most closely related work in adaptation definitions and code generation using machine learning. Finally, in \fref{sec:conclusion} we discuss current shortcomings of Auto-\ac{COP} and provide directions how to address them in future work.


\section{Background}
\label{sec:background}

Auto-\ac{COP} enhances the \ac{COP} paradigm with a learning 
technique to generated adaptations from previous observed executions. 
Before diving into the details of our proposal, this section 
introduces the main concepts of \ac{COP}, and \ac{RL} options, the 
learning technique used to extend \ac{COP}. To illustrate both 
concepts, we use examples from the driving assistant system. 
	
\subsection{\acl{COP}}
\label{sec:cop}

The \ac{COP} paradigm proposes a programatic technique to 
dynamically adapt the system's behavior to the context, in a highly 
modular and reusable fashion~\cite{hirschfeld+08cop}. \ac{COP} 
languages provide specialized abstractions fostering the definition of 
dynamic adaptations as system modules that are independent from each 
other, and the base system, but yet are highly composable.

\ac{COP} languages introduce adaptations as the combination of two 
abstractions, \emph{contexts} and \emph{behavioral variations}. 
Contexts represent situations from the execution environment 
captured by system variables (\ie the state) that are 
monitored internally, or can be sensed by external sensing devices. 
Whenever specific environment conditions for context objects are 
satisfied, the contexts are said to be \emph{active}; otherwise they 
are \emph{inactive}.
Each context is associated with a set of behavioral variations 
specifying adaptations to the normal behavior specified by the base 
system. As contexts become active, their associated behavioral 
variations are made available in the system --that is, this will be the 
observed system behavior. Internally, upon context activation, its 
associated behavioral variations are composed with the system at run 
time. Context deactivation withdraws all behavioral variations 
associated with the context from the system at run time. 
In both cases, the system behavior is effectively adapted.

We use Context Traits\footnote{Available at \url{https://www.npmjs.com/package/context-traits}}~\cite{gonzalez+13ctxtraits}, a 
\ac{COP} extension of ECMAScript, as the implementation language for 
our proposed approach. However, note that this work is not specific to 
this language, and is applicable to \ac{COP} in general. To make 
explicit the concepts introduced by \ac{COP}, in the following we show 
the main \ac{COP} constructs and their interplay using Context Traits. 
Constructs defined in other \ac{COP} languages are conceptually 
similar, although syntactic differences may 
exist~\cite{salvaneschi+12survey, appeltauer+09cop}.
	
Programmatically, contexts are first-class system entities (\ie objects) that abstract 
semantically meaningful situations gathered from the surrounding 
environment (or internal state) of the system, using sensors and monitors.
For example, in our driving assistant, a \ctx{closeProximity} 
context is defined to represent the situation in which a vehicle 
is in close proximity in front (\eg $v2$ in \fref{fig:overtake}), as 
shown in \fref{lst:ctx-definition1}. 

\begin{ctxtraits}[label={lst:ctx-definition1},
	frame=lines,
	caption={Static definition of a context object}]
 closeProximity = new cop.Context({
    name: "vehicle in close proximity"
 })
\end{ctxtraits}

The context defined in \fref{lst:ctx-definition1} is abstracted from 
the information gathered from the environment variables (sensors).
In our example, the \ctx{closeProximity} context is associated with 
the readings from the proximity sensor. \fref{lst:ctx-variable} shows 
the conditions, over the system state, in which the context should be active.

\begin{ctxtraits}[frame=lines,
	label={lst:ctx-variable},
	caption={Context activation conditions}]
 if(proximitySensor.receive() < 300)
   closeProximity.activate()
 else
   closeProximity.deactivate() 
\end{ctxtraits}	

Whenever a sensed variable satisfies a specified condition, the 
context defined by the variable is \emph{activated} using the 
\ac{COP} \scode{activate()} construct. If the condition is no longer 
satisfied, the context is \emph{deactivated} using the \ac{COP}
\scode{deactivate()} construct. 
As contexts are activated, their associated behavioral variations (explained next) are 
composed with the running system, enabling the behavior they define, 
and overwriting/extending the behavior previously defined. 
As contexts are deactivated, the system is recomposed not to 
take into account the behavioral variations associated with the 
context.

Behavioral variations are first-class system modules, providing 
fine-grained partial behavior definitions of system entities (\eg new 
states, or behavior specializations). For example, the specialized 
behavior to manage proximity to a vehicle can be defined as 
in \fref{lst:behavioral-adaptation1}, effectively adapting the base 
\scode{drive} behavior (continue \scode{straight()}) defined for the vehicle, 
with new actions to execute instead (\ie \scode{steerLeft()} and \scode{steerRight()}). 
As explained before, 
the vehicle should steer to avoid the slow vehicle in front. The 
sensed \ctx{closeProximity} context is associated with its behavioral 
variations by means of the \scode{adapt} construct, as shown by 
\fref{ln:adapt} of \fref{lst:behavioral-adaptation1}.

Putting contexts, behavioral variations, and context activation together, 
we adapt the behavior of a program at run time by the dynamic recomposition 
of the system behavior upon context activations.

\begin{ctxtraits}[frame=lines,
	numbers=left,
	label={lst:behavioral-adaptation1},
	caption={Statically defined behavioral adaptation to avoid slow vehicles in front}]
 closeProximityBehavior = Trait({
    drive: function() {
       steerLeft()
       steerRight()
    }
 })
 closeProximity.adapt(vehicle, closeProximityBehavior) ` \label{ln:adapt} `
\end{ctxtraits}

Behavioral variations define sequences of actions (\ie functions) to 
execute whenever a particular context is sensed. In Auto-COP, we 
automate adaptation definition given a context and a sequence of 
actions, similar to the proposal of \ac{RL} option learning, which we 
present next. 

\subsection{Reinforcement Learning Options}
\label{sec:reinforcement-learning}

Temporally extended actions (\eg macro actions) are used in Artificial 
Intelligence applications to ensure robustness and build in prior 
knowledge into applications, from early work on building and executing 
robot plans~\cite{fikes1993} to recent applications in deep 
learning~\cite{zhang2019}.  In particular, in \ac{RL}, macro-actions 
are used to speed up learning, or to minimize the periods of 
suboptimal performance during exploratory interaction with the 
environment. We consider \ac{RL}-based macro-action techniques, called 
options~\cite{sutton98intra}, to be suitable for learning and 
integrating sequences of actions into \ac{COP}-based adaptive systems. 
Adaptations in \ac{COP} respond to changes in the context, akin to 
the way actions in \ac{RL} are learned and taken in response to 
observed environment conditions. 

\ac{RL} is a learning technique that learns optimal actions for 
specific environment conditions by trial-and-error based on 
interactions with the environment. At each timestep, an \ac{RL} agent 
perceives the environment and maps it to a state $s_i$ from its state 
space $S$. It then selects an action $a_i$ from its action set $A$ 
and executes it. The agent receives a reward $r_i$ from the 
environment when it transitions to the next state, based on which it 
updates the suitability of taking the action $a_i$ in state $s_i$. 
The agent's goal is to learn a policy (\ie the most suitable action 
for each of its states) to maximize the long-term cumulative reward. 
The learning rate $\alpha$ determines to what extent new experience 
overwrites previously learned ones, and the discount factor $\gamma$ 
determines how much are the future rewards discounted, in order for 
agents to prioritize immediate actions, but still be able to plan the 
best longterm actions. 

Early interactions with the environment are focused on environment 
exploration, \ie actions are picked randomly and uniformly from 
actions available in a given state, while later stages, after an 
agent has had a chance to learn the quality of actions, are focused on 
exploitation --that is, executing mostly those actions known to lead 
to the highest long-term rewards. An \ac{RL} option encodes sequences 
of such actions into temporally-extended actions, and is defined by three 
components: a policy $\pi$, which is mapping from a set of states $S$ 
to a set of actions $A$, an initiating condition (or a set of conditions) 
$I \subseteq S$, and a terminating condition which determines its length. 
Numerous ways to build options from atomic actions exist, from 
those that are manual~\cite{elfwing04} or require domain 
knowledge~\cite{mcgovern98}, to learning options~\cite{randlov98}, 
identifying sequences that lead to the fulfillment of 
subgoals~\cite{stolle02}, or automatically generating all action
combinations but using another layer of learning to narrow down the 
full set of generated macro-actions to the most useful 
ones~\cite{pickett02}. Options can also be learned from generated 
behavior histories, by extracting the most commonly used sequences of 
atomic actions~\cite{girgin05}. 
Option learning in Auto-\ac{COP} is most closely related to the 
techniques used in  subgoal fulfillment~\cite{stolle02}, and behavior 
histories~\cite{girgin05}.


\section{Auto-COP Design}
\label{sec:learning-adaptations}
 
This section presents the design of Auto-\ac{COP}, our approach for 
the automated generation of adaptations for self-adaptive systems. We 
first present the high-level functionality of Auto-\ac{COP}, and then 
describe its two main steps: 
\begin{enumerate*}[label=(\arabic*)]
\item learning action sequences using \ac{RL} options, and 
\item automating the generation of \ac{COP}-based adaptations using 
the learned options. 
\end{enumerate*}

\subsection{Adaptation Generation in Auto-\ac{COP}}
\label{sec:auto-cop}

In \ac{COP}, behavioral variations are predefined by developers, 
explicitly specifying: the adaptation behavior, the base objects 
affected by the adaptation, the context in which the adaptation takes 
place, and the conditions for the selection and scoping of such 
context. However, this information may not be available or known by 
developers beforehand. This restricts the adaptivity of the system to 
the known and specified behavior. Furthermore, it is difficult to know 
whether the predefined adaptions are indeed the most appropriate 
behavior for the situation in the current environment, as the 
environment can evolve during the execution.

To address these problems, Auto-\ac{COP} enables the 
dynamic generation of adaptations based on previous system executions, 
and interaction with the environment. 
 
\begin{figure}[htbp]
  \centering
  \includegraphics[width=\columnwidth]{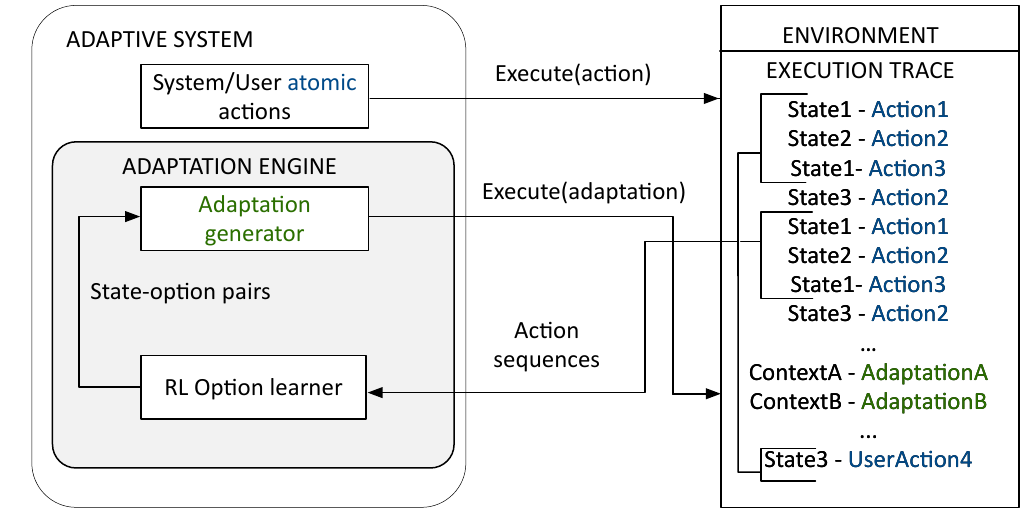}
  \caption{Adaptation learning process model}
  \label{fig:process}
\end{figure} 
  
The overall process of automated adaptation generation using 
Auto-\ac{COP} is shown in \fref{fig:process}. Software systems 
execute sequences of actions, generating an execution trace of 
actions. The source of these actions may be predefined behavior, user 
intervention (which correspond to atomic actions - in blue), or generated 
by other processes (which correspond to 
adaptations - in green). The trace of executed actions is used 
as input for the \ac{RL} 
\emph{Option learner}, which learns the most suitable action 
sequences for specific environment conditions. These sequences are 
then used as input of the \emph{Adaptation generator}, which packages 
them into reusable \ac{COP} adaptations, by defining a context 
object from the system state, and the behavior variations associated 
with such context from the learned options. 

\subsection{Learning Options}
\label{sec:MatchingBehaviorsToContexts}

The process to learn options continuously executes throughout the 
system's lifetime. The process consists of generating adaptations 
from the best suitable options extracted from the system's execution 
trace of actions (both atomic actions, and learned options). This process is 
described using the pseudo code in 
\fref{lst:autoCOPruntime}. 
To learn options, we observe the set of environment states $S$ (\ie 
contexts) accessible by the system (\eg monitored variables), and 
the set of (atomic) actions $A$ that can execute for 
each state (\ie functions). The \emph{Option learner} module learns a 
set of options $O$, which contains a map of all states $s_i \in S$, 
associated to a set of actions sequences, defined as options $O_i$, 
available for execution in that state. All option sets $O_i$ are 
initially empty, as options are learned during the execution. 
 
\begin{ctxtraits}[label={lst:autoCOPruntime},
	frame=lines, numbers=left,
	caption={Auto-COP run-time adaptation loop}]
 `$S := \{s_1, ..., s_n\}$ `
 `$A := \{a_1, ..., a_m\}$ ` 
 `$O := \{\{s_1, \emptyset\}, ...,  \{s_n, \emptyset\}\}$ `
 lastBatchEnd := 0 
 
//OPTION EXTRACTION
while(true) { 
  AtomicActionLog:=write(`$s_i, a_i, r(s_i,a_i)$`) `\label{ln:option-init}`
  if(timestep `$t \bmod batchSize$ == $0$`) { 
    //build options from batches in the execution trace (Log)
   for(i=lastBatchEnd; i<logSize; i++){
     //read  the state from the Log
     `$loggedState$` := readLogLine(i, statePosition) 
     for(j=0; j<maxOptionLength && !goalReached; j++){
       //read action from the Log
       `$actionSequence_i$` += readLogLine(i+j, actionPosition) 
     }
   `$O_i$` := {`$loggedState, actionSequence_i$`}
   O.add(`$O_i$`)
   }
   lastBatchEnd := logSize
  } `\label{ln:option-end}`

  //ADAPTATION GENERATION
  ReinforcementLearning.initialize(`$S, O$`) `\label{ln:main-init}`
  currentState := senseEnvironment()
  //epsilon-greedy option selection 
  if(contextAdaptationAvailable(currentState)) { 
    selectedOption := currentState.pickAnOption(`$\epsilon$`)
    currentState.activate()
    execute(selectedOption)  //execute the adaptation
    currentState.deactivate()
    ReinforcementLearning.updateOption(`$S, O, r(currentState, selectedOption)$`)
  } else { //no options available, execute atomic action
    atomicAction := currentState.pickAction()
    execute(atomicAction) 
  }
  newCOPAdaptation := generate(currentState, highestQOption)
  O.add(newCOPAdaptation) `\label{ln:main-end}` 	
 }
\end{ctxtraits}
 
During the early execution steps, only atomic actions (\ie 
predefined 
or human-enacted) are executed as the system has no custom adaptations 
built from options yet. The execution of each action is logged 
in the execution trace, together with the state (\ie values of known variables) 
in which it executed 
(Execution Trace to the right of~\fref{fig:process}). For each action, 
we also log the reward based on the effects of that action on the 
current state. This 
reward, $r(s_i,a_i)$, can have multiple sources. For example, if the 
underlying actions are learned using a learning-based system, the 
reward corresponds to the reward obtained from the environment. 
If actions are enacted by a human, a 
fixed positive reward can be associated with each action under the 
assumption that such interventions are performed by qualified system 
operators. Finally, a constant reward (\eg 1) can be given each time 
a state-action pair is encountered, under the assumption that the
actions more frequently executed are those that are more suitable. 

Our algorithm first processes the execution trace in the \ac{RL} 
\emph{Option learner} module to populate the options sets $O_i$ in 
small batches, every $batchSize$ number of steps. We opt for batch 
processing, as processing after every execution step can be costly, 
but have a negligible effect on the options generated. The details of 
the process correspond to the Lines 
\ref{ln:option-init}-\ref{ln:option-end} of 
\fref{lst:autoCOPruntime} (\scode{OPTION EXTRACTION}). 
 
The outcome of this process is $O$, the map of all states identified 
in the execution trace, each associated to a set of option sequences 
ranging in length from 1 (\ie a single action) to the maximum option 
length $n$. The maximum option length can be either specified 
externally, calculated experimentally taking into account the 
frequency and usefulness of recorded sequences, or set to the 
maximum number of actions required to reach the system's goal 
state. To illustrate the generated output, \fref{lst:ordered} shows a 
generic example of possible action sequences for the states 
\scode{stateVariablesSet1} and \scode{stateVariablesSet2}. 
 
\begin{ctxtraits}[frame=lines,
 	numbers=left,
 	label={lst:ordered},
 	caption={Ordered action sequences extract}]
 	state: [stateVariablesSet1]
 	reward:1 -> actions: ["action3"]
 	reward:4 -> actions: ["action3",    "action1"]
  	reward:10 -> actions: ["action3",   "action1","adaptation1"]
 	reward:10 -> actions: ["action3",   "action1", "adaptation1", "action4"]
 	
 	state: [stateVariablesSet2]
 	reward:11 -> actions: ["action1"]
 	reward:17 -> actions: ["action1",   "adaptation1"]
\end{ctxtraits}
 
Note that each state may contain multiple extracted options, and the 
sequences in these options may include previously generated options 
(shown as \scode{"adaptation"} in the snippet), following
our example. However, most of the extracted options will be 
unsuitable for adaptation, as they may not yield a correct system 
state. Auto-\ac{COP} narrows down options, to select a single option 
as the most appropriate behavior for each state (context), which we 
use to generate a \ac{COP} adaptation. We describe this process next. 
 
\subsection{Automated Adaptation Generation}
\label{sec:packaging}
 
The Auto-\ac{COP} \emph{Adaptation generator} module takes as input 
the different options for each explored state and uses an \ac{RL} 
process to learn the most appropriate option to execute as an 
adaptation. The generation process is specified in Lines 
\ref{ln:main-init}-\ref{ln:main-end} of \fref{lst:autoCOPruntime} 
(\scode{ADAPTATION GENERATION}). 
For each option executed at a given \scode{currentState} we record 
the reward for executing the option using standard Q-learning (state, 
action, reward) updates, aiming to maximize long-term system 
performance. After sufficient exploration, the options with the 
highest Q-values (\ie highest expected long-term rewards) for each 
state are identified. The reward model for options takes into account 
the system reward for reaching the final state of the option. We use 
this model as we are interested in the system to reach its final goal, 
only those options closing the gap from the current state to the goal 
state are considered as more appropriate.
The options with the highest reward are the ones used in the 
\emph{Adaptation generator} module, to generate the context and 
behavioral variation objects.   
 
\begin{ctxtraits}[frame=lines,
 	numbers=left,
 	label={lst:behavior-generation},
 	caption={Adaptation generation stub}]
 ContextCurrentState = new cop.Context({name: "currentState" }) ` \label{ln:ctx-gen} `
 BehavioralVariation = Trait({ ` \label{ln:ba-init} `
    option : function() { 
      //learned action sequence 
       action`$_1$`();
       `$\ldots$`
       action`$_n$`();
    }
 }); ` \label{ln:ba-end} `
 ContextCurrentState.adapt(BaseSystem, BehavioralVariation); ` \label{ln:ctx-assoication}  `
\end{ctxtraits} 
 
The definition of generated contexts is shown in \fref{ln:ctx-gen} of 
\fref{lst:behavior-generation}. Each context is given as a name the 
string literal corresponding to the state in which it must take place. 
The generation of behavioral variations constitutes the re-definition 
of the system's base behavior, by using the sequence of actions of the 
selected option, as shown in Lines \ref{ln:ba-init}-\ref{ln:ba-end} of 
\fref{lst:behavior-generation}. Finally, we also generate the 
association of behavioral variations with their respective context; 
this is done in \fref{ln:ctx-assoication} in 
\fref{lst:behavior-generation}.

Once adaptations are generated, while the system further executes, 
whenever the context associated with one of the adaptations is 
sensed, the system activates the corresponding context, 
\scode{ContextCurrentState.activate()}. This triggers the composition 
of the behavioral variation associated with the context with the 
system to be executed. Once the behavioral 
variation executes, the context is deactivated, 
\scode{ContextCurrentState.deactivate()}, and the system reverts to
its base behavior executing atomic actions. 

This process continuously executes during the lifetime of the system: 
traces are logged, processed into options in batches, and their 
suitability explored in interaction with the environment to propagate 
the most suitable options into \ac{COP} adaptations. As the system 
executes, executed options can be taken into account for the 
generation of new options, effectively composing options within 
options. Additionally, note that, if 
and when the environment conditions change, the appropriateness of 
adaptation may change, new options can receive a better reward than 
the options used to generate the current adaptations (\eg adaptations 
get a negative reward, or different atomic actions will be executed 
by the system users). Such process enables the continuous generation 
of adaptations. In this way Auto-\ac{COP} removes the need for 
adaptations to be predefined at design time and ensures the system 
can continuously adapt as conditions change without manual changes to 
the source code. In the next section we illustrate applications of 
Auto-\ac{COP} in two case studies, and evaluate its performance in 
satisfactorily generating adaptations that meet specified system 
goals.


\section{Validation}
\label{sec:validation}

This section shows the feasibility of Auto-\ac{COP} as a means to 
automate adaptation generation in self-adaptive systems. To 
validate our approach, we present two different case studies in 
which we validate our adaptation generation technique by exploring 
different system characteristics and application domains. The 
objective of these case studies is to evaluate the correctness of 
generated behavior 
(\ie it does not lead to errors, and is beneficial towards the system 
goal, measured according to application-specific metrics), and the 
usefulness of generated adaptations in the system (as opposed to 
executing only atomic actions).

\subsection{Driving Assistant}
\label{sec:driving-assistant}

Our first case study consists of a driving assistant system, with the 
purpose of assisting drivers while driving in a two lane highway, as 
briefly introduced in the motivation in the introduction 
(\fref{sec:introduction}). The objective of this case study is to 
evaluate the correct generation of adaptations from observed behavior 
(\ie it does not lead to errors).

The expected system 
behavior is to drive in the righthand side lane, and only use 
the left lane to overtake, if the vehicle in front is driving too 
slowly. The vehicle has three goals: to drive at or under the speed 
limit, avoid crashing into other vehicles in 
front of it, and avoid driving in the left lane.  We use the vehicle 
behavior with respect to these goals to measure the system's 
performance. The evaluation metrics correspond to the number of times 
the vehicle crashes, the number of times the vehicle is in the wrong lane, 
and the number of times the vehicle drives over the speed limit. All three 
metrics should be minimized.

\subsubsection{Environment}
\label{sec:assistant-environment}
For the driving assistant scenario, we developed a small simulation 
environment in JavaScript.\footnote{Available at: \url{https://github.com/FLAGlab/DrivingAssistant}}

The system's environment consists of a two lane road of $500km$. The  
road speed limit is set to $60km/h$, with other traffic vehicles 
appearing with a $10\%$ chance, at least 3 time steps after the last 
vehicle is left behind. Traffic vehicles drive at a speed of $30km/h$ 
(so that they can be overtaken) and always appear on the driving lane 
(\eg the righthand lane in the environment). The vehicle's atomic actions 
are to \scode{speedUp()} and \scode{slowDown()} (each modifying the 
current speed by $\pm 10km/h$ from the current speed), going
\scode{straight()} to maintain the current speed, \scode{steerLeft()} 
to move from the right lane to the left lane, and \scode{steerRight()} 
to move from the left lane to the right lane. 

All the experiments were executed locally on a MacOSX system using 
version v18.8.0 of Node.js.

\subsubsection{Execution}
\label{sec:assistant-execution}

To generate a log of atomic actions for our driving 
assistant to learn from, we 
develop an \ac{RL} algorithm that learns the correct driving behavior 
for the vehicle from the atomic actions. Note that  
Auto-\ac{COP} is agnostic with respect to the origin of such
actions in the execution trace (\eg actions could be generated 
by human interventions, but as manually controlling a vehicle for the 
duration required for significant evaluation is cumbersome, we  
automated the atomic action generation using \ac{RL}). In addition, 
by using \ac{RL} to generate atomic actions, the system 
occasionally executes an erroneous action (when it is exploring), 
therefore allowing us to illustrate how option generation corrects 
for those actions by generating adaptations based on their impact on 
the system execution rather than just merely repeating execution 
traces. 

In the (atomic) action learning process, the state space of the 
driving assistant is given by the vehicle's speed, taking discrete 
values multiple of $10km/h$ in the interval $[0, 70]$, the current 
driving lane, modeled as $0$ for driving on the right lane and $1$ 
for driving on the left lane, and the proximity to the car in front, 
divided in the discrete interval $[1,4]$, describing the time steps 
to reach the vehicle
in front. A proximity of 4 denotes that there is no vehicle in front, 
while values 1, 2, and 3 denote there is a vehicle 
in immediate proximity in front, and a crash is to occur in 1, 2, or 3 
timesteps into the future if no action is taken. The reward model 
penalizes crashing ($-8$), driving in the wrong lane ($-5$), driving 
over the speed limit ($-6$), driving too slowly ($-6$), and provides a 
positive reward when driving in the right lane with no car in front 
($8$). The learning rate $\alpha$ is set 
to 0.1, the discount factor $\gamma$ to 0.6, and action selection is 
$\epsilon$-greedy, with $\epsilon$ starting at 0.2 in the exploration 
stage, reducing to 0.001 in the exploitation stage. 

We let the training \ac{RL} system run for $8000$ steps, generating an 
execution trace of $8000$ atomic actions. In the execution trace, 
for each action we record the current system state 
$[speed, lane, vehicle\_proximity]$, the (atomic) action executed, 
the next state after executing the action, and the reward obtained by 
executing said action. 

After the execution traces are recorded, we execute the 
\scode{OPTION EXTRACTION} stage, Lines 
\ref{ln:option-init}-\ref{ln:option-end} of 
\fref{lst:autoCOPruntime}, to build possible options (action 
sequences). Based on the execution trace, in this step we generated a 
total of 26 options for execution in 13 different states. Only one 
option is selected for each state to generate the appropriate 
adaptation. Note that there are 72 total states in the environment, 
but some of the states do not have any associated options in 
consequence of two reasons: some states are not experienced during 
exploration (\ie do not appear in the execution 
trace), and we are only interested in generating options for the 
states in which adaptations are required (\ie only when system's goals 
are not met). If the vehicle is driving at the correct speed limit 
(state value $60km/h$), in the correct lane (state value $0$), and 
there is no vehicle in front (state value $4$), then no adaptation is 
required. 

\begin{table*}[htbp]
	\centering
	\caption{State space for the learned options and the corresponding generated adaptations in the driving assistant case study}
	\label{tab:assistant-state-space}

\begin{tabular}{ c c c | l | p{7.9cm} | l }   \hline 
\toprule
\multicolumn{3}{c |}{\textbf{State}} & \textbf{No.} & \textbf{Action sequence} & \textbf{Executions}\\
\midrule

  \textbf{50} & \textbf{0} & \textbf{1} & \textbf{1} & \textbf{\{steerLeft(), speedUp(), steerRight()\}} & \textbf{29}\\
\hline
  50 & 0 & 1 & 2 & \{steerLeft(), speedUp(), steerRight(), steerLeft(), steerRight()\}  & 3\\
  \hline
  \textbf{60} & \textbf{0} & \textbf{1} & \textbf{1} & \textbf{\{steerLeft(), steerRight()\}}  & \textbf{656}\\
\hline
  60 & 0 & 1 & 2 & \{steerLeft(), steerLeft(), steerRight()\}  & 1\\
\hline
  60 & 0 & 1 & 3 & \{slowDown(), speedUp(), speedUp(), speedUp(), straight(), speedUp(), speedUp(), speedUp()\} & 1\\
\hline
  60 & 0 & 1 & 4 & \{steerLeft(), straight(), steerRight(), steerLeft(), steerRight()\} & 1\\
  
\bottomrule
\end{tabular}

\end{table*}

To illustrate this process we focus on the example of the 
overtaking behavior (not pre-defined as an atomic action) when a
traffic vehicle appears in front. \fref{tab:assistant-state-space} 
shows all the extracted options for two of such states. Intuitively, 
overtaking a vehicle in front is done by a maneuver sequence 
\{\scode{steerLeft()}, \scode{steerRight()}\} of atomic actions. 
Therefore, the desired adaptation in the state \scode{[50,0,1]} is 
option number 1 in the table, as it overtakes the vehicle in front, 
while also speeding up to the target speed of $60km/h$, to end up in 
the goal state \scode{[60,0,4].} Similarly, in the state 
\scode{[60,0,1]} the desired behavior is option 1, to overtake a 
vehicle without any additional actions. Indeed, the frequency of the 
execution of these actions in the original trace file shows these to 
be the most executed options for these states. However, the table also 
shows additional options generated for such states. These options were  
executed a single time during exploration. While we cannot 
match these traces to an exact situation during the execution, we 
speculate that option 2 in state \scode{[50,0,1]} and option 4 in 
state \scode{[60,0,1]} are a result of the vehicle encountering 
a traffic vehicle in its lane immediately after overtaking a first 
vehicle, resulting in having to repeat the \{\scode{steerLeft()}, 
\scode{steerRight()}\} maneuver one more time. For option 3 in the 
state \scode{[60,0,1]} we speculate that the first incorrect action 
in the sequence (\scode{slowDown()}) resulted in the crash of the 
vehicle, consequently yielding a speed of $0$, from which the 
vehicle had to execute \scode{speedUp()} 6 times to reach the target 
state \scode{[60,0,4]}. The process of adaptation generation should, 
however, be able to identify these sequences as not suitable, and 
learn to extract option 1 into a \ac{COP} adaptation, for each of our 
example states. 

The list of extracted options, states and their associated action 
sequences, serves as the input to the next stage of the learning 
process, where we explore options further using \ac{RL} to identify 
which option is the most suitable for the system. To do this, we 
execute the \scode{ADAPTATION GENERATION} stage, Lines 
\ref{ln:main-init}-\ref{ln:main-end} of \fref{lst:autoCOPruntime}. 
To generate adaptations we use the same state space, rewards, and 
learning parameters as specified for the \ac{RL} process 
above. However, in this case, we enable the use of all available 
options for each state, together with the 5 atomic actions. The 
best option (\ie the option with the highest reward) is selected to 
generate a  \ac{COP} adaptation for the given state. We evaluate and 
present the performance of the system in the exploitation stage, 
comparing the performance of just using the atomic actions and 
using the Auto-\ac{COP} combinations of atomic actions and 
generated adaptations. 

\begin{figure}[htbp]
	\centering
	\begin{subfigure}[t]{\linewidth}
		\includegraphics[width=\columnwidth]{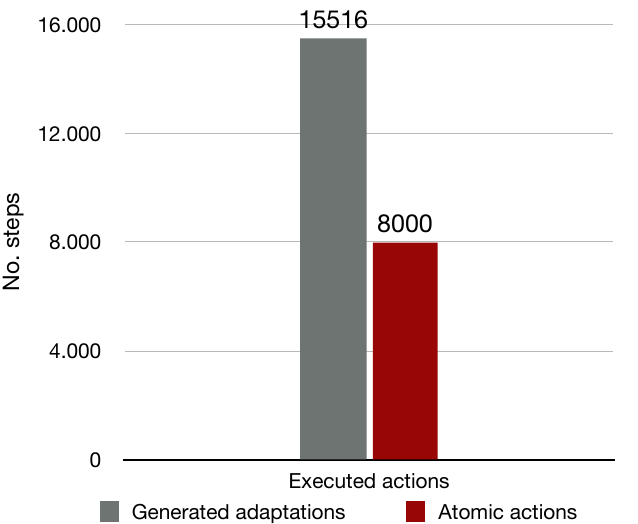}  
		\caption{Number of executed actions}
		\label{fig:assistant-steps}
	\end{subfigure} \\
	\begin{subfigure}[t]{\linewidth}
		\includegraphics[width=\columnwidth]{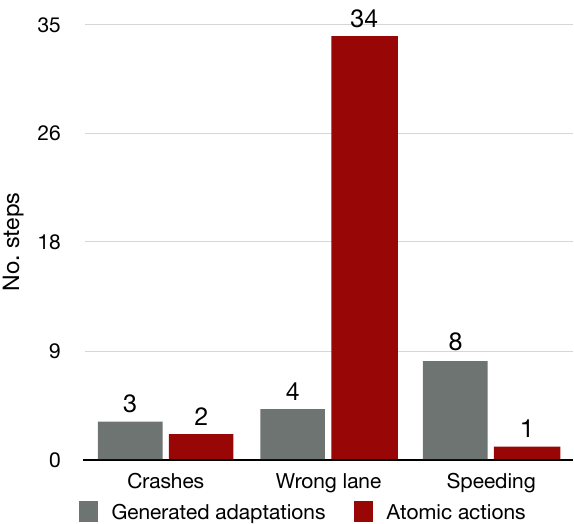}
		\caption{Correct behavior}
		\label{fig:assistant-behavior}
	\end{subfigure}
	\caption{Results of behavior correctness and usefulness of generated adaptations for the driving assistant}
	\label{fig:assistant-results}
\end{figure}

\fref{fig:assistant-results} shows the effectiveness of the generated 
adaptations as measured by the performance with respect to the three 
system goals, comparing the atomic actions implementation and the 
Auto-\ac{COP} generated 
adaptations. 
First we compare the decision points between the two systems, 
running each of them for 8000 execution steps. In accordance with the use of options, the 
less decision points the agent takes, the faster is going to be its executions. In our case, 
since we are executing for 8000 decision points in both scenarios, the efficiency of the agent 
is given by the amount of atomic actions it executes in those 8000 decision points. The more 
atomic actions executed, the more adaptations are used by the agent, showing their appropriateness.
\fref{fig:assistant-steps} shows the amount 
of actions executed by the vehicle when using the generated 
adaptations vs. using only atomic actions. We actuate a total of 
1992 adaptations for the 8000 execution steps (\ie decision points), 
resulting in a total 
of 15516 executed actions vs the 8000 atomic actions (as only a 
single atomic action can be executed per decision point in a time-step). 
Therefore, we confirm that the use of options generates suitable 
sequences of actions, and that the system learns to use those as 
adaptations to successfully enhance the system behavior, almost 
doubling the efficiency of system execution, or 
halving the number of system interventions required. 

Second we compare the correctness of using Auto-\ac{COP} vs. using 
atomic actions. \fref{fig:assistant-behavior} shows the 
correctness of the resulting behavior generated by Auto-\ac{COP}, as 
the number of rule violations incurred by the vehicle, the fewer violations, 
the more correct is the approach. We observe that 
there are far less lane violations when using adaptations (4 wrong 
lanes) vs. using atomic actions (34 wrong lanes). We also observe 
that the vehicle presents an additional crash event (3 crashes) when 
compared to the atomic actions execution (2 crashes). This can be 
explained from two perspectives. 
First, using adaptations, we effectively execute almost twice as many 
steps that in the case of atomic actions, encountering more than 
double the amount of cars in traffic (1497 vs. 692), therefore while 
the total amount of crashes increases, its percentage decreases by 
$0.25\%$. This constitutes and improvement over the base system 
behavior.
Second, upon further inspection of the execution trace, we note that 
two of the crashes take place during the execution of atomic 
actions, rather than as a consequence of executing the adaptations (as 
noted previously, the final behavior of the system implementing 
adaptations is a combination of atomic actions and adaptations, as 
not all states have adaptations associated with them). Therefore the 
amount of crashes when using the generated adaptations is further 
reduced to $50\%$. Finally, we observe that the number of speed-limit 
violations increased with respect to the case using atomic actions, 
going over the speed limit 8 times, vs. 1 speed-limit violation. 
However, on balance, we can conclude that the use of generated 
adaptations is beneficial for the system performance, as the overall 
number of violations across the three metrics significantly decreases. 

The adaptations generated by Auto-\ac{COP} for the two discussed 
states \scode{[60,0,1]} and \scode{[50,0,1]} are shown in 
\fref{lst:Context6001} and \fref{lst:Context5001}. In both cases the  
behavioral variations generated (\scode{BAContext6001} and 
\scode{BAContext5001}) correctly match the desired overtake behavior.  

\begin{ctxtraits}[frame=lines,
	numbers=left,
	label={lst:Context6001},
	caption={Generated adaptation for state \scode{[60,0,1]}}]	
  Context6001 = new cop.Context({ name: "Context6001"})
  BAContext6001 = Trait({ 
		option: function(){ 
			this.steerLeft();
			this.steerRight(); 	
		}
  })
  Context6001.adapt(agent, BAContext6001)
\end{ctxtraits} 

\begin{ctxtraits}[frame=lines,
	numbers=left,
	label={lst:Context5001},
	caption={Generated adaptation for state \scode{[50,0,1]}}]
  Context5001 = new cop.Context({ name: "Context5001"})
  BAContext5001 = Trait({ 
		option: function(){ 
			this.steerLeft();
			this.speedUp();
			this.steerRight(); 	
		}
  })
  Context5001.adapt(agent, BAContext5001)
\end{ctxtraits} 

To integrate these adaptations, we use \ac{COP} in the driving 
assistant application effectively adapting the application's behavior 
from using atomic actions to using the generated behavioral 
variations whenever appropriate.
\fref{lst:cop-integration} shows the definition of the base driving 
agent (Lines \ref{ln:agent-init}-\ref{ln:agent-end}), together with 
the overall driving assistant behavior, in which, for each step, 
depending on the current state, we choose an action to execute (Lines 
\ref{ln:state-init}-\ref{ln:state-end}). If the current state does not 
have an associated Auto-\ac{COP} adaptation, we execute a primitive 
action. If the state has an associated adaptation to it, then we 
execute the correpsponding behavioral variation in the \scode{option} 
function. To execute the adaptation we follow the process for dynamic 
adaptations used in \ac{COP}, as explained in \fref{sec:cop}. First, 
we activate the context representing the current state 
(\fref{ln:context}). This composes the behavioral variation associated 
to the context with the system. In our example, for the state 
\scode{[60,0,1]}, the behavioral variation in \fref{lst:Context6001} 
is composed with the system. The effect of this composition is 
that the \scode{Agent} now has an \scode{option} function defined. 
Second, we use the behavioral variation by calling the generated 
option (\fref{ln:option}). Finally, the system transitions to a 
different state as a consequence of the option execution, and the 
context is deactivated (\fref{ln:deactivate}).

\begin{ctxtraits}[frame=lines,
	numbers=left,
	label={lst:cop-integration},
	caption={Integration of generated adaptations using \ac{COP}}]
class Agent { `\label{ln:agent-init}`
	speedUp() { ... }
	slowDown() { ... }
	steerRight() { ... }
	steerLeft() { ... }
	straight() { ... }
}	`\label{ln:agent-end}`
		
while(true) {
 if(qtable[this.currentState]) `\label{ln:state-init}`
   action = qtable[this.currentState]
 else  
   action = this.randomAtomicAction() `\label{ln:state-end}`
 if(action >= Agent.actions.length) {
   eval('Context`\$`{state}'.activate()) `\label{ln:context}`
   agent.option()	`\label{ln:option}`
   eval('Context`\$`{state}'.deactivate()) `\label{ln:deactivate}`
 } else 
   eval('agent.`\$`{actions[action]}'())
}
\end{ctxtraits}

\subsection{Warehouse Delivery Robot}
\label{sec:delivery-robot}

The second case study focuses on evaluating the correct generation of 
adaptations and the usefulness of such adaptations. This case study 
consists of a robot delivery system for a 
warehouse. The purpose of the robot is to move packages from their 
storage space in a warehouse to the collection point at the front 
desk. The robot's goal is to pick-up a package ready for 
delivery and take it to the front desk. While the robot does not 
know where a package is located, once it picks it up, it must 
take it to the front desk at a defined goal location as specified by 
a coordinate pair $(x, y).$\footnote{This example is inspired by the standard RL benchmark Taxi-v3.}
The base system behavior for the robot is to roam the warehouse 
unaware of the packages or the drop-off point. As requests for a 
package come in, then the robot should go and fetch them. Fetching 
each of the products in the warehouse then constitutes a special 
situation, the robot should adapt its behavior to.

\subsubsection{Environment}
\label{sec:robot-environment}

The system environment is delimited by a $n\times n$ grid ($5\times 5$ in our evaluation).\footnote{Available at: \url{https://github.com/FLAGlab/RobotDelivery}} 
The packages are located at specific positions 
within the grid, different from the point at the front desk. 

The robot can move to any of the adjacent 
positions from its current position $(x, y)$, within the boundaries 
of the environment. Action \scode{north()} is used to move to cell 
$(x-1, y)$, \scode{south()} to move to cell $(x+1, y)$, 
\scode{east()} to move to cell $(x, y+1)$, and \scode{west()} to 
move to cell $(x, y-1)$. Additionally, the robot has the 
\scode{pickup()} and \scode{dropoff()} actions to pick-up 
and drop-off the package respectively, once it is at the correct 
location.

\subsubsection{Execution}
\label{sec:robot-execution}

Following the same process as in the driving assistant case study, we 
define an \ac{RL} algorithm that learns the correct behavior to 
pick-up and and drop-off packages using atomic actions. The robot's 
state space is defined by its $(x, y$) coordinates (with values 
in $[0, 4$]), and a single boolean value denoting whether the robot 
is available for package pickup. The reward model for the robot gives 
a positive reward ($20$) for correctly dropping-off the package and 
($10$) for correctly picking it up, and a negative reward ($-10$) for 
executing an incorrect action (pick-up or drop-off at a wrong 
location, picking-up when there is already a package, or dropping-off 
without a package), all other actions give a small negative reward 
($-1$). The learning rate $\alpha$ is set to 0.1, the discount factor 
$\gamma$ to 0.6, and action selection is $\epsilon$-greedy, with $
\epsilon$ starting at 0.2 in the exploration stage, reducing to 0.001 
in the exploitation stage. 

Unlike the driving assistant which executes continuously without a 
terminal state, the warehouse delivery robot is an episodic task 
--that is, the task ends when the package is delivered to the desired 
destination. Therefore, the execution is organized into episodes. 

To create the log of atomic actions from which we extract the 
adaptations, we first train an \ac{RL} agent for 600 episodes 
(\ie 600 instances of picking-up and dropping-off a package) 
to explore the state space and find correct routes. In the execution trace we 
record the action taken at each step, the current robot's state 
$[pos\_x, pos\_y, package?]$, the next state after executing the action, and the reward obtained 
from the action execution. 

After the traces are recorded, we execute the option building stage, 
Lines \ref{ln:option-init}-\ref{ln:option-end} in 
\fref{lst:autoCOPruntime}, to build possible options. Based on the 
execution trace, in this step we generated a total of 65 options for 
execution in 11 states (out of total of 50 states). The next step is 
to reduce available options to the option with the highest Q-value to 
generate the adaptation, using the adaptation generation stage (Lines 
\ref{ln:main-init}-\ref{ln:main-end} in \fref{lst:autoCOPruntime}). 
For this we use the same state space, rewards, and learning parameters 
as specified for the \ac{RL} process above, while the action set now 
consists of both the existing atomic actions and all of the 
generated options. 

\begin{figure}
	\centering	
	\includegraphics[width=1.01\columnwidth]{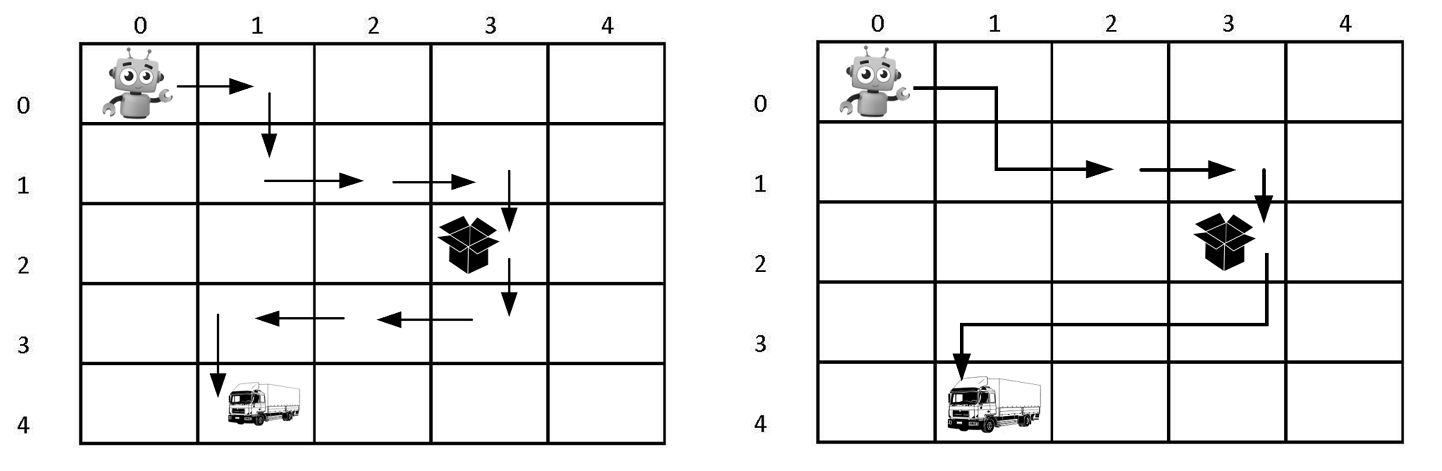}
	\caption{Delivery robot's path with atomic actions and Auto-\ac{COP} adaptations}
	\label{fig:robot-behavior}	
\end{figure}

We compare the performance of the system using the atomic actions 
and the Auto-\ac{COP} generated adaptations, during 600 exploitation 
episodes (we use the same number of episodes as for the RL-only training stage, for ease of comparison with atomic actions). There are no dropoffs to incorrect locations in either the 
atomic actions case nor the adaptations case, so the performance of 
Auto-\ac{COP} preserves correct behavior with respect to the system 
goal. As there is only a single goal, and is always met in both cases, 
we do not show graphs for this case study. Instead, we illustrate the 
resulting behavior, \ie the path that the warehouse robot took, while 
executing atomic actions (left-side of \fref{fig:robot-behavior}) 
vs. the path executing adaptations (right-side of 
\fref{fig:robot-behavior}). 

In the case of atomic actions, a total of 11 actions execute to 
achieve the goal from the starting location: 9 steps that move the 
robot around the grid to correct locations, 1 pickup, and 1 dropoff 
action. In the case of adaptations, a total of only 5 executions took 
place: an option consisting of step sequence $\{east, south, east\}$, 
3 atomic actions $east$, $south$, $pickup$ and the final option 
that took the robot all the way from the pickup location to the 
delivery location, consisting of sequence $\{south, west,$ $west, 
south, dropoff\}$. The generated \ac{COP} code for this adaptation is 
shown in \fref{lst:Context23false}. Therefore, we conclude that 
Auto-\ac{COP} was successfully applied in the warehouse delivery robot 
system, as it identified correct action sequences to package into 
adaptations, resulting in meeting the goal in each episode, and 
requiring only 4 execution steps to achieve the goal vs. 11 steps 
required for atomic actions. 

\begin{ctxtraits}[frame=lines,
numbers=left,
label={lst:Context23false},
caption={Generated adaptation for state \scode{[2,3,false]}}]
  Context23false = new cop.Context({ name: "Context23false"})
  BAContext23false = Trait({   
	 option: function() { 		
	 	 this.south();		
	 	 this.west();		
	 	 this.west();		
	 	 this.south();		
	 	 this.dropoff(); 	
 	}
  })
  Context23false.adapt(agent, BAContext23false)
\end{ctxtraits} 

The integration of the generated adaptation in 
\fref{lst:Context23false} with the warehouse delivery robot 
application follows a similar process and structure to the one 
explained in the case of the driving assistant.


\section{Related Work}
\label{sec:related}

This sections discusses Auto-\ac{COP} in the perspective of 
existing related approaches to define adaptations, applications of RL in adaptive systems, as well as code 
generation techniques using \ac{ML}.

\subsection{Adaptation Definition}

Different techniques have been proposed for the definition of 
adaptations in self-adaptive systems. From the programming language 
perspective, existing approaches are based on 
\ac{COP}. As previously mentioned, all existing 
approaches~\cite{appeltauer+09cop, salvaneschi+12survey} require the 
upfront definition of adaptations by developers.

From a modeling and architecture perspective, adaptations follow the  
MAPE feedback loop, in which it is the developers' 
responsibility to model and define all system components to be able 
to monitor the environment information, analyze it, and process and 
execute the appropriate behavior from a set of available adaptations. 
Adaptations are introduced into the system using predefined 
hook-points in the base application. While the MAPE model is most 
commonly used for the definition of self-adaptive systems, other 
models such as feedback control loops, the self-management model, or 
the autonomic computing reference architecture share the property of 
requiring the upfront definition of adaptations~\cite{villegas17}.

Advances in big data technology have led to the use of data streams 
coming from large cyber physical systems and smart city systems to drive the 
adaptation of their behavior. High-rate data processing enables the 
definition of an architecture to adapt application behavior based on 
big data analysis, as presented by RTX~\cite{schmid17}. In RTX, 
applications knobs are defined to change the system behavior based 
the analysis of processed data. Unlike Auto-\ac{COP}, RTX requires 
the definition of the conditions of the analyzed data in order to 
enact changes. This is similar to the definition of adaptations (and 
their conditions), which we do not require in Auto-\ac{COP}.

Adaptations are managed and defined in the domain of software 
variability by means of variability models realized using \acp{SPL}. 
Magus~\cite{bashari18} uses dynamic \acp{SPL} for the adaptation of 
web service composition in the case one of the service components 
fails. The Magus approach uses a context state model, that must exist 
beforehand, containing the functional and non-functional properties of 
services. This model is then used to modify or adapt the service 
properties when changes are detected. However, this model can only 
adapt and react to those functional and non-functional aspects already 
defined for the web services.

Recently, \ac{ML} approaches have been explored for the definition of 
adaptations as a means to overcome manual system modeling, or 
adjusting the models as they evolve~\cite{jamshidi19}, similar to the 
objective of Auto-\ac{COP}. \citet{jamshidi19} present an \ac{ML} 
algorithm used to learn optimal configurations for a planning robot. 
The configurations from which the robot learns are given beforehand, 
and compared with the robots's plan (\ie objective) to find the most 
appropriate configuration for a given state. The configuration is 
able to change (through online learning) based on the robot's state. 
While the online learning follows a similar process to the one 
proposed in Auto-COP, the robot actions (\ie configuration) are 
predefined, unlike the actions we extract for our adaptations, which 
are learned from previous interactions with the environment.

\subsection{Reinforcement Learning in Adaptive Systems}

\ac{RL} is extensively used in (self-)adaptive systems due to its ability to learn 
suitable system behavior, requiring only interaction with the environment, without 
an up-front model of that environment, allowing adaptation of the system as 
conditions change at runtime. Some of the early \ac{RL} approaches use in load 
balancing and resource allocation were  presented in a manifesto on \ac{RL} 
applications in autonomic computing~\cite{tesauro07}, expanding from there to 
complex multi-agent scenarios. A recent survey~\cite{dangelo19} found that over 
half learning-based complex adaptive systems rely on \ac{RL} for implementation 
of adaptive behavior. The applications are varied, ranging from large-scale 
cyber-physical infrastructure (\eg transportation~\cite{dusparic12taas} and smart 
grids~\cite{marinescu17}), service composition~\cite{caporuscio2016, wang19}, internet 
of things~\cite{restuccia20}, information systems~\cite{palm2020}, and architectural 
adaptation~\cite{filho2022}, to name a few, and cover both classical tabular RL 
approaches and neural network-based Deep RL approaches. 
However, unlike in \ac{COP} adaptations, \ac{RL}-based adaptive behavior is 
intertwined with the application code, which could lead to decreased code 
quality~\cite{cardozo23prevelance}, decreased maintainability of the codebase and 
increased technical debt, similar to ML-based applications in general~\cite{sculley14}. 
By combining adaptivity of \ac{RL} with modularity and reusability of \ac{COP}, 
Auto-\ac{COP} combines to the benefits of the two approaches, maintaining \ac{RL} 
ability to adapt online, while reducing entanglement between core codebase and new 
adaptations. Auto-\ac{COP} does not necessarily replace the existing \ac{RL}-based 
approaches, but can build on top; underlying atomic actions can be generated using 
existing \ac{RL}  approaches, whether tabular or deep, following which Auto-\ac{COP} 
can, using \ac{RL} options built from generated actions, package them into adaptations.

\subsection{Code Generation Using \acl{ML}}

Auto-\ac{COP} utilizes \ac{RL} to generate \ac{COP} adaptations at 
run time and therefore can be classified as an 
\ac{ML}-based approach for code generation. 

The area of code generation using \ac{ML} is an emerging field due to 
recent advancement in deep learning techniques and their increased 
applicability to different areas. 
A full review of the wider field of deep learning 
for software engineering and future research directions is presented 
by~\citet{devanbu2020deep}, while~\citet{cruz-benito21} provide a 
review and comparison of deep learning approaches to automated source 
code generation and auto-completion. Applications of learning in the 
code generation focus on, for example, predicting a sequence of 
source code tokens~\cite{tiwang2019}, generating code based on 
natural language code descriptions~\cite{sun2019}, or generating 
front end code from hand-drawn wireframes~\cite{sharma2020}. The 
only work that utilizes \ac{RL} in this field focuses on source code 
summarization to generate code comments~\cite{wang2020}. To the best 
of our knowledge, no approach currently utilizes \ac{RL} options, 
nor focuses on generating methods out of sequences of existing 
behavior, as Auto-\ac{COP} does.


\section{Conclusion and Future Work}
\label{sec:conclusion}

This paper presents Auto-\ac{COP}, a new technique for the automated  
generation of adaptations for self-adaptive systems realized using 
\ac{COP}. Our proposal extends the state-of-the-art in \ac{COP} with 
an online adaptation engine based on \ac{RL} options. Auto-\ac{COP} 
enables the system to learn the behavior of the adaptations for
specific system states, rather than solely relying on predefined 
adaptations by developers. The online adaptation engine continuously 
processes execution traces to extract action sequences from the set 
of atomic actions executed for every state. Taking the 
extracted state-action sequences (\ie options), the system learns  
the most appropriate option, from a set of available extracted options, 
with respect to the system goal. Learned options are used to generate 
behavioral adaptations associated with the context in which they 
should take place. Generated behavioral variations are used to adapt 
the base-system behavior, whenever their associated context is sensed 
in the environment.

Our approach is evaluated using systems from two different 
application domains. In both experiments, we evaluated the 
effectiveness of Auto-\ac{COP} in generating and exploiting generated 
adaptations. Our experiments show that for both systems, first, the 
most appropriate learned options generated adaptations from previous 
execution, and second, adaptations are effective in contributing to 
meet the system goal, whenever their context is activated.

One of advantages of Auto-\ac{COP} is that it enables learning of 
adaptations at run time, while retaining the modularity aspects 
offered by \ac{COP}. Additionally, Auto-\ac{COP} is fully compatible 
with existing \ac{COP} tools and techniques. Future 
work should address the integration with such techniques, to enable 
a more robust context-adaptation approach. For example, conflict 
between adaptations can occur if multiple contexts are sensed 
simultaneously, requiring execution of multiple adaptations. Such 
conflicting interactions can be resolved using a W-learning 
technique~\cite{cardozo17cop}, which determines the adaptation with 
a higher priority, to execute at that particular system state. 
Similar techniques could be integrated with Auto-\ac{COP} to compose 
adaptations whenever multiple adaptations are sensed 
simultaneously~\cite{cardozo20comina}. 

As generated adaptations become more complex and consist of longer 
sequences, the possibility of environment changes mid-adaptation 
increases. In this situation, long sequences of actions being executed 
without intermediate system state checking would result in inappropriate 
or even dangerous actions being executed, if
adaptations are always executed to completion once triggered. 
Monitoring for environment changes continuously during adaptation execution and enabling 
adaptation interruptions is required to guarantee safe system 
performance. Integration with human-in-the-loop systems should also 
be investigated, to enable users or operators to override adaptations 
by means of manual execution of atomic actions when interrupting 
an adaptation. Such intervention/override can be taken into account 
in the option learning process as a negative feedback, which will 
lead to updating the adaptation based on atomic actions enacted 
by the human. This should also be done in conjunction with more 
fine-grained approaches to integrating \ac{RL} and human feedback 
(RLHF approaches); currently human actions are "taken for granted", 
\ie it is assumed that skilled human operator actions are correct. In 
reality, those actions can sometimes be inconsistent or even wrong, 
so integration of Auto-\ac{COP} with \ac{RL}-based techniques which 
are robust to this inconsistency, such as policy shaping~\cite{Griffith2013}, 
can be investigated. 

The applicability of this work to other areas of adaptive systems 
should also be investigated. For example, Self-healing 
systems~\cite{ghosh07} provide the capability 
for systems to diagnose and recover from errors. These approaches 
use diagnosis components (\eg root cause analysis~\cite{sama10}) to 
identify the cause of failures, and the access points (locations) in 
which a sequence of recovery actions can be used to assure the 
continuity of the system. The overall functioning of self-healing 
systems maps to our proposed approach to generate adaptations, 
understanding adaptations as recovery actions and contexts as the 
root cause of the failure. Given this mapping, we could learn 
recovery actions as they take place, and then automate them so that 
future occurrences of the failure would not require any intervention.

\bibliographystyle{elsarticle-num-names}
\biboptions{sort&compress}
\bibliography{local,bib/general,bib/compsci,bib/learning}

\end{document}

\endinput